\documentclass[acus]{JAC2000}
\usepackage{epsfig,amssymb}    

\def\3{\ss}
\newcommand{\be}{\begin{equation}}
\newcommand{\ee}{\end{equation}}
\newcommand{\bea}{\begin{eqnarray}}
\newcommand{\eea}{\end{eqnarray}}

\usepackage{graphicx}

\begin{document}
\title{
DIPOLE MODE DETUNING IN THE NLC INJECTOR LINACS
\protect\thanks{Work supported by the U.S.\ Department
of Energy under contract DE-AC03-76SF00515.
}\vspace*{-2mm} }

\author{K.L.F.\ Bane, Z. Li, SLAC, Stanford University,
 Stanford, CA 94309, U.S.A.}
\maketitle

\vspace*{-2cm}
\section{Introduction}\label{sec:intro}

A major consideration in the design of the accelerator structures in the
 injector linacs of the JLC/NLC\cite{zdr}
is to keep the wakefield effects within tolerances for both the nominal
 (2.8~ns)
and alternate (1.4~ns) bunch spacings.
One important multi-bunch wakefield effect 
is  beam break-up (BBU),
where a jitter in injection conditions of a
bunch train is amplified in the linac;
another is static emittance
growth caused by structure misalignments.

The injector linacs 
comprise the prelinac, the $e^+$ drive linac,
the $e^-$ booster, and the $e^+$ booster. The first three will
operate at S-band, the last one, at L-band.
Compared to the
main (X--band) linac, the wakes will tend to be smaller by a factor 
$1/64$ and $1/512$, respectively, for 
 the S-- and L--band linacs.
This reduction, however,---especially for the S-band machines---,
 by itself, is not sufficient.
Two ways of reducing the wake effects further are
to detune the first pass-band dipole modes
 and to damp them.
In this report our goal is to design the accelerator structures for the
injector linacs using detuning alone,
an option that is simpler than including damping.
We will consider only the
effects of modes in the first dipole pass-band, whose
strengths overwhelmingly dominate. The effects of the higher
pass-band modes, however, will need to be addressed in the future.
For a more detailed version of this work see Ref.~\cite{BL}.
Note that the design of the $e^+$ booster structure, which is straightforward,
will not be discussed here.

Machine properties for the injector linacs are given in 
Table~\ref{taone}.
Shown are the initial and final energies $E_0$, $E_f$, the
machine length $L$, the initial (vertical) beta function
averaged over a lattice cell
$\bar\beta_{0}$, and the parameter $\zeta$ for a rough fitting of the beta
function to $\bar\beta\sim E^\zeta$.
The rf frequencies are sub--harmonics of 
11.424~GHz.
As for beam properties, for the nominal
bunch train configuration (95 bunches spaced at 2.8~ns),
the particles per bunch $N=1.20$, 1.45, 1.45, $1.60\times10^{10}$
and normalized emittance $\epsilon_{yn}=3\times10^{-8}$,
$10^{-4}$, $10^{-4}$, .06~rm, for
the prelinac, e$^+$ drive, e$^-$ booster, and e$^+$ booster, respectively.
For the alternate configuration (190 bunches spaced at 1.4~ns)
$N$ is reduced by $1/\sqrt{2}$.

\begin{table}[htb]
\caption{Machine properties of the injector linacs.}
\vskip1mm
\centering
\label{taone}
\begin{tabular}{||l||c|c|c|c||} \hline \hline
Name   &$E_0$, $E_f$[GeV] & $L$[m]  &  \raisebox{-.2ex}[0pt]{$\bar\beta_{0}$[m]} &  $\zeta$   \\ \hline\hline
Prelinac &    1.98, 10.0     & 558    &   8.6               &    1/2      \\ \hline
$e^+$ Drive &  0.08, 6.00     & 508    &   2.4               &    1/2      \\ \hline
$e^-$ Booster &  0.08, 2.00     & 163    &   3.4               &    1/4       \\ \hline
$e^+$ Booster & 0.25, 2.00     & 184    &   1.5               &     1       \\ \hline\hline
\end{tabular}
\end {table}

\vspace*{-1.7cm}
\section{Emittance Growth}

\subsection{Beam Break-Up (BBU)}
In analogy to {\it single-bunch} BBU in a linac\cite{chao},
multi-bunch BBU can also be characterized by a strength
parameter, but one dependent on bunch number $m$:  
\begin{equation}
\Upsilon_m= {e^2NLS_{m}\bar\beta_0\over 2E_0} 
g(E_f/E_0,\zeta)\quad\quad[m=1,\ldots,M]\
,\label{eqeta}
\end{equation}
with $M$ the number
of bunches in a train.
The sum wake 
\begin{equation}
S_m= \sum_{i=1}^{m-1} W[(m-i)\Delta t]\quad\quad\quad[m=1,\ldots,M]\ ,
\end{equation}
with $W$ the transverse wakefield
and $\Delta t$ the time interval between bunches in a train.
The wake, in turn, is given by a sum over the dipole modes
in the accelerator structures:
\begin{equation}
W(t)= \sum^{N_m}_n 2 k_{n}\sin({2\pi f_{n}t/ c})
\exp(-\pi{f_{n}}t/Q_n)\quad,
\label{eqwake}
\end{equation}
with $t$ time and
$N_m$ the number of modes; $f_n$, $k_n$, and $Q_n$ are, respectively,
the frequency, the kick factor, and the quality factor of the $n^{\rm th}$
mode.
The function $g(x)$ in Eq.~\ref{eqeta} depends on the
focusing profile in the linac.
Assuming the beta function varies as $\bar\beta\sim E^\zeta$,
\begin{equation}
g(x,\zeta)= {1\over\zeta}\left({x^\zeta-1\over x-1}\right)
\quad\quad\quad[{\bar\beta\sim E^\zeta}].
\end{equation}

If $\Upsilon_m$, for all $m$, is not large, the linear approximation
applies, and this parameter
directly gives
 the (normalized) growth in amplitude of bunch $m$.
The projected (normalized) emittance growth of the bunch train then
becomes (assuming, for simplicity,  that, in phase space, the beam ellipse
is initially upright) 
$\delta\epsilon\approx {1\over2}\Upsilon_{rms0}^2y_0^2/\sigma_{y0}^2$,
with $\Upsilon_{rms0}$
the rms with respect to 0 of the strength parameter,
$y_0$ the initial bunch offset, 
and $\sigma_{y0}$ the initial beam size.
As jitter tolerance parameter, $r_t$, we can take that ratio $y_0/\sigma_{y0}$
that yields a tolerable emittance growth, $\delta_{\epsilon t}$.

\vspace*{-3mm}
\subsection{Misalignments}
If the structures in the linac are (statically) misaligned with respect
to a straight line, the beam at the end will have an 
increased projected emittance.
If we have an ensemble of misaligned linacs then,
to first order, the 
distribution in emittance growth at the end of these linacs
is given by an exponential distribution
$\exp[-\delta\epsilon/\langle\delta\epsilon\rangle]/\langle\delta\epsilon\rangle$, 
with\cite{static}
\begin{equation}
\sqrt{\langle\delta\epsilon\rangle}= {e^2NL_a(x_a)_{rms}{S}_{rms}\over E_0}
\sqrt{{N_a{\bar\beta_0}\over2}}\,h(E_f/E_0,\zeta) 
\label{eqmisa}
\end{equation}
with $L_a$ the structure length, 
$(x_a)_{rms}$ the rms of the structure misalignments,
${S}_{rms}$ the rms of the 
sum wake {\it with respect to the average}, and
$N_a$ the number of structures;
the function $h$ is given by
(again assuming $\bar\beta\sim E^\zeta$):
\begin{equation}
h(x,\zeta)= \sqrt{{1\over\zeta x}\left({x^\zeta-1\over x-1}\right)}
\quad\quad\quad[{\bar\beta\sim E^\zeta}].
\end{equation}
Eq.~\ref{eqmisa} is valid assuming the so-called betratron term
in the equation of motion is small compared to the misalignment term.
We can define a misalignment tolerance:  
$x_{at}=(x_a)_{rms}\sqrt{\delta\epsilon_t/\langle\delta\epsilon\rangle}$,
 with $\delta\epsilon_t$ the tolerance in emittance growth.

We are also interested
in the tolerance to cell-to-cell misalignments
caused by fabrication errors. A structure is built
as a collection of cups, one for each cell, that is brazed together,
and there will be errors, small compared
to the cell dimensions, in the straightness of each structure.
To generate
a wake (for a beam on-axis) in a structure 
with cell-to-cell misalignments
we use a perturbation approach based on the
 eigenmodes of the
unperturbed structure\cite{perturb}\cite{BL}. 

\vspace*{-3mm}
\section{Wakefield Design}
Reducing emittance growth requires reducing the sum wake.
In the main (X-band) linac of the NLC, 
the strategy to do this is to use
Gaussian detuning to 
generate a fast Gaussian fall-off in the wakefield envelope; in particular,
at the position of the second bunch the wake is reduced by roughly
2 orders of magnitude from its initial value.
At the lower frequencies of the injector linacs we have fewer
oscillations between bunches and this strategy requires too much detuning.
Instead, we will follow a
 strategy that puts early bunches on zero crossings of the
wake, by a proper choice of the average frequency.
As for the distribution of mode frequencies,
we will aim for a uniform
distribution, for which
the wake is (for $\pi\bar{f}t/Q$ small):
\begin{equation}
W\approx {2\bar{k}\over N_m}
\sin(2\pi{\bar f}t)\,{{\rm sin}(\pi{\bar f}t\Delta_{\delta f})\over
  {\rm sin}(\pi{\bar f}t\Delta_{\delta f}/N_m)}\quad,\label{equni}  
\end{equation}
with $N_m$ the number of modes, $\bar k$ the average kick factor,
$\bar f$ the average frequency, and
$\Delta_{\delta f}$ the full width of the distribution.
The wake envelope initially drops with $t$ as a sinc function,
but eventually resurges again, to a maximum 
at $t=N_m/({\bar{f}\Delta_{\delta f}})$.

For the 2nd bunch to sit on the zero crossing requires that
$\bar{f}\Delta t=n/2$, with $n$ an integer.
For S-band, given our implementation of the SLED-I pulse
compression system, the optimal rf efficiency
is obtained when the average dipole mode frequency
is 4.012~GHz. 
For this case,
with the alternate (1.4~ns) bunch spacing,
$\bar{f}\Delta t=5.62$.
The half-integer is 
achieved by changing ${\bar f}$ by $-2\%$, a change
which, however, results in a net loss
of 7\% in accelerating gradient.
One way of avoiding this loss is to reduce the group velocity by
increasing
the phase advance per cell of the fundamental mode
from the nominal $2\pi/3$.
In fact, we find that by going to $3\pi/4$ phase advance
we can recapture this loss in gradient.

For the resurgence in the wake to occur after the 
bunch train has passed requires that $\Delta_{\delta f}$ 
be significantly less than $N_m/(M{\bar f}\Delta t)$, which, 
in our case, is about 10\%.
Another possibility for pushing the resurgence 
to larger $t$ is to use two
structure types, which can effectively double the number of modes available for
detuning.
This idea has been studied; 
it has been rejected in that it requires tight alignment
tolerances between pairs of such structures.

\vspace*{-3mm}
\subsection{Optimization}
The cells in a structure are coupled to each other, and
to obtain the wakefield
we need to solve for the eigenmodes of the system. 
We obtain these numerically
using a double-band circuit model~\cite{Gluck}. 
The computer program we use generates $2N_c$
coupled mode frequencies $f_{n}$ and kick factors $k_{n}$,
with $N_c$ the number of cells in a structure.
It assumes the modes are trapped at the ends of the structure.
We will use only the first $N_c$ modes (those
of the first pass-band) for our wakefield
since they overwhelmingly dominate
and since those of the second band are not obtained accurately.

\begin{figure}[htbp]
\centering
\epsfig{file=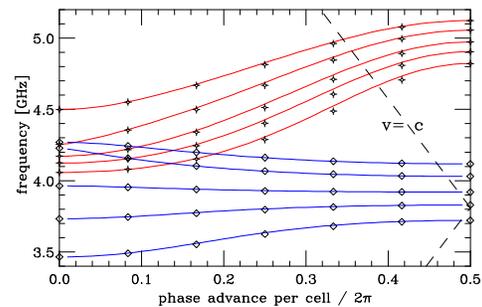, width=62.mm}
\caption{
The dispersion curves of the first two dipole bands
of representative cells in a
$3\pi/4$ structure. 
}
\label{fiomegbet_3pi4}
\end{figure}

The constants (circuit elements) for the program are obtained
by fitting to 
results of a 2D electromagnetic program OMEGA2\cite{OMEGA2}
applied to representative cell geometries, and then using interpolation.
Here we consider structures of the disk--loaded type, 
with rounded irises.
The iris and cavity
radii are adjusted to give the correct fundamental mode frequency
and the desired synchronous
dipole mode frequency.
Therefore, cell $m$ can be specified by one free
parameter, the synchronous frequency (of the first
dipole mode pass-band). 
The $3\pi/4$ S-band structure
consists of 102 cells with a cell period of 3.94~cm,
iris thickness of 0.584~cm, and cavity radius $\sim4.2$~cm;
the $Q$ due to wall losses (copper) $\sim14,500$.
Fig.~\ref{fiomegbet_3pi4} shows the first two dispersion curves 
of representative cell geometries
(for iris radii from 1.30 to 2.00~cm). 
The plotting symbols give the OMEGA2 results,
the curves, those of the circuit program.

\vspace*{-3mm}
We will consider a uniform input (synchronous)
frequency distribution, but with
a slanting top.
This leaves us with
3 parameters to vary:
the (relative) shift in average frequency (from a nominal 4.012~GHz)
$\delta\bar{f}$, the (relative) width of the distribution $\Delta_{\delta f}$,
and the tilt parameter $\alpha$
($-1\le\alpha\le1$, with $\alpha=1$ giving a right triangle
distribution with positive slope).
Varying these parameters
we calculate $S_{rms0}$ and $S_{rms}$
for the coupled modes, and
for both bunch train configurations, and we optimize.
We find that a fairly optimal case consists of
$\delta\bar{f}=-2.3\%$, $\Delta_{\delta f}=5.8\%$, and $\alpha=-0.20$,
where $S_{rms0}=S_{rms}=.004$~MV/nC/m$^2$.
In Fig.~2 we show the dependence of $S_{rms0}$ on
$\delta\bar{f}$ and $\Delta_{\delta f}$ near the optimum.

\begin{figure}[htb]
\vspace*{-1mm}
\epsfig{file=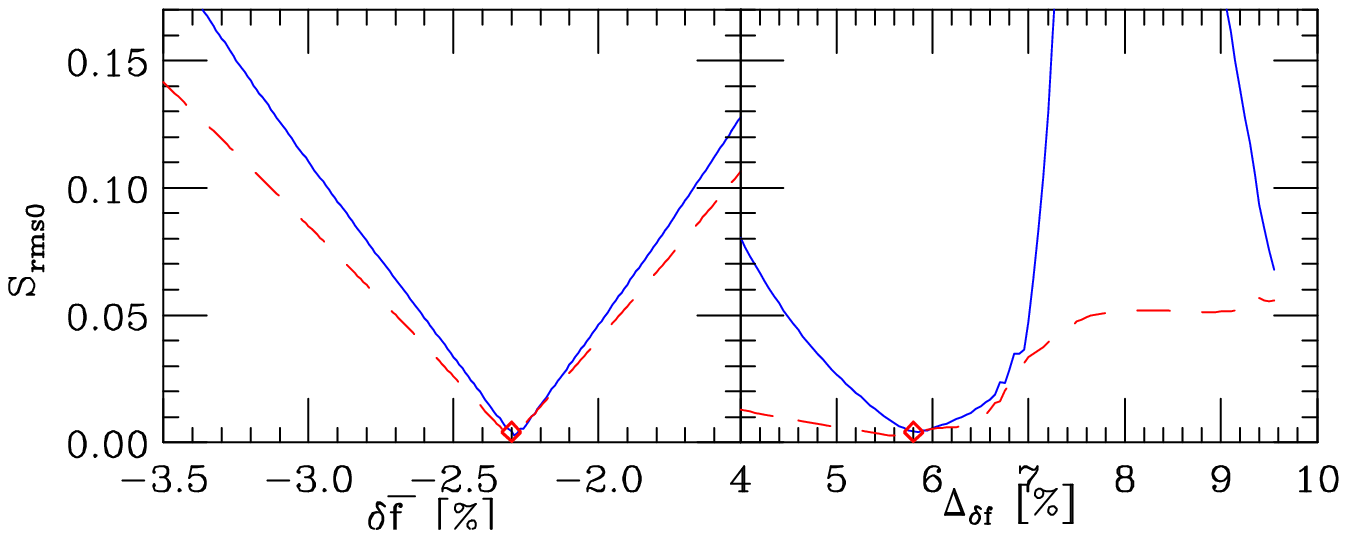, width=77.5mm}
Figure 2: $S_{rms0}$
[MV/nC/m$^2$] {\it vs.} $\delta {\bar f}$ and $\Delta_{\delta f}$ 
near optimum, for
$\Delta t=2.8$~ns (solid) and 1.4~ns (dashes).
\stepcounter{figure}
\label{fiopt}
\end{figure}

\vspace*{-3mm}
In Fig.~3 we display, for the optimal case, 
the frequency distribution~(a), the kick factors~(b), and the envelope of the wake~(c).
The dashed curves in (a) and (b) give the synchronous (input) values.
The plotting symbols in (c) give $|W|$ at the bunch positions for
the alternate (1.4~ns) bunch train configuration. 
In (b) we see no spikes, thanks to the 
fact that the synchronous point is near pi, and,
serendipitously, 
$f_0<f_\pi$ for cell geometries near the beginning of
the structure,
$f_0>f_\pi$ for those near the end\cite{Gluck}.
(Note that for the optimized $2\pi/3$ structure, for which 
$f_0>f_\pi$ for all cell geometries, 
there is such a spike, and consequently $S_{rms0}$ is 5 times larger
than here\cite{BL}.)
From (c) we note that many of the earlier bunches 
have wakes with amplitudes significantly
below the wake envelope.

\begin{figure}[htbp]
\centering
\vspace*{-1mm}
\epsfig{file=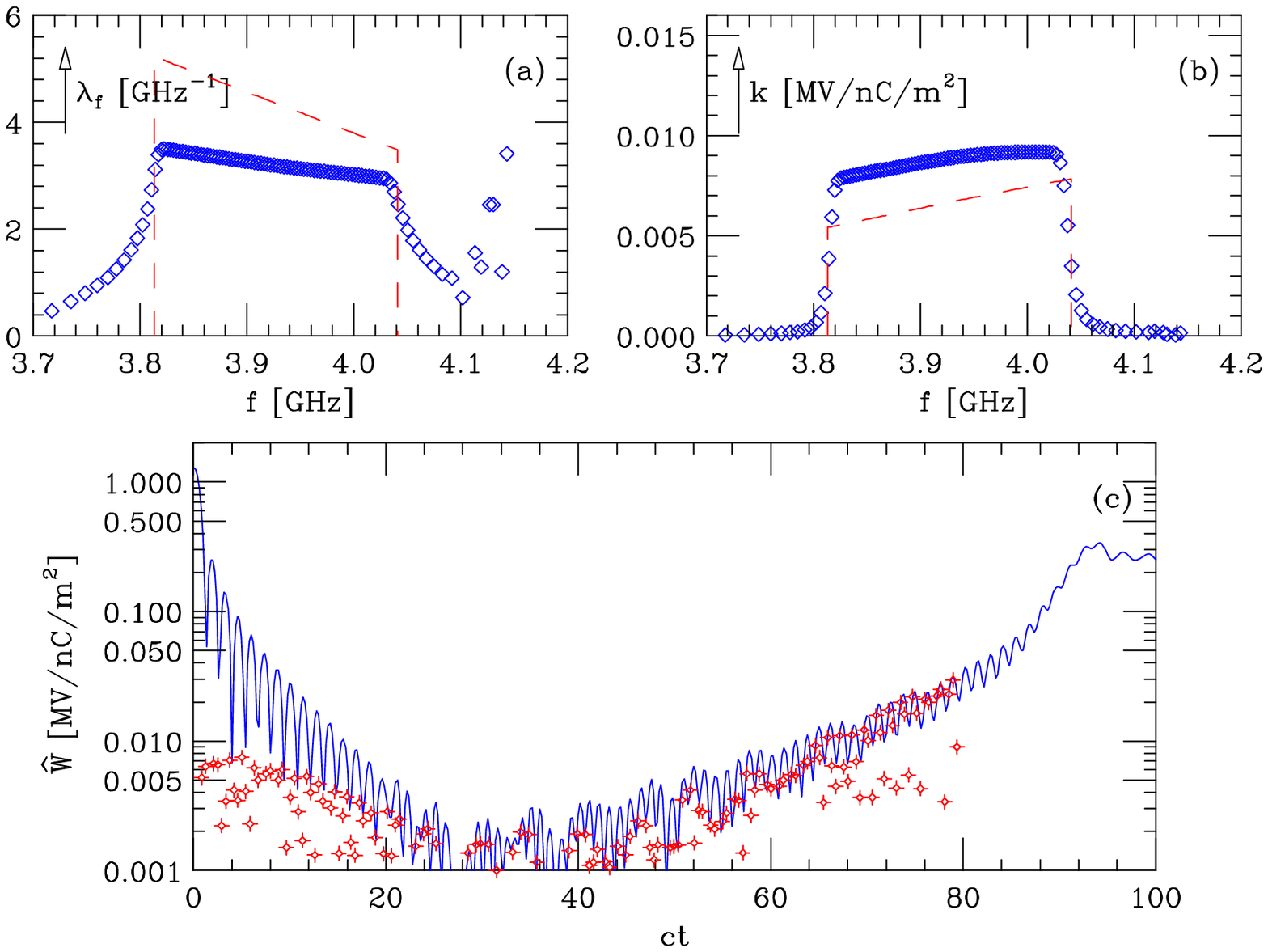, width=77.5mm}
Figure 3: Results for the optimal $3\pi/4$ structure.
\stepcounter{figure}
\label{fisband_opt_3p4}
\vspace*{-3mm}
\end{figure}

\vspace*{-4mm}
\subsection{Frequency Errors}
\begin{figure}[htb]
\centering
\vspace*{-1mm}
\epsfig{file=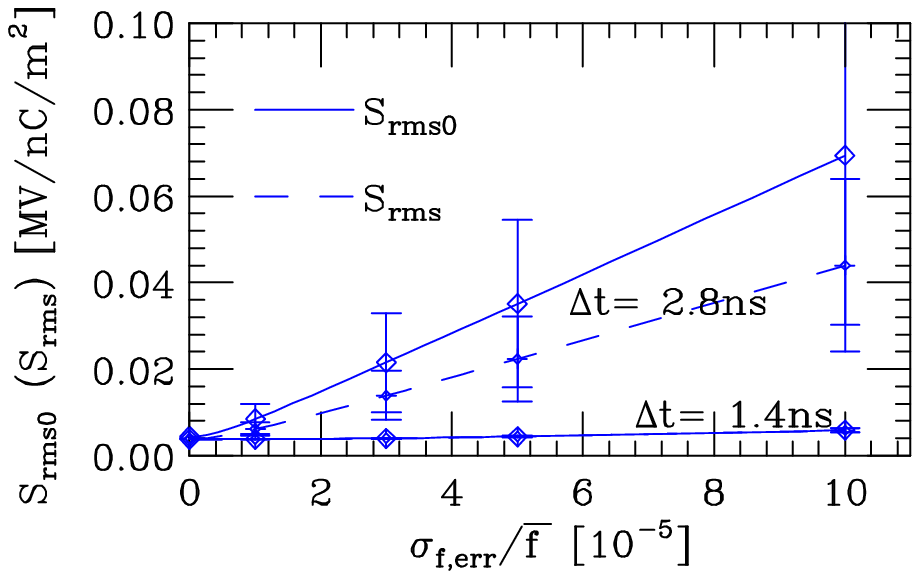, height=36mm}
\caption{
The effect of random frequency errors.
}
\label{fierrs}
\end{figure}

Errors in cell manufacturing will result in frequency errors.
In Fig.~\ref{fierrs} we give $S_{rms0}$
and $S_{rms}$, when a
random error component is added to the (input) synchronous frequencies
of the optimal distribution (each plotting symbol,
with its error bars, represents 400 seeds).
With a frequency spacing of $\sim8\times10^{-4}$, an rms frequency error
of $1\times10^{-4}$ is a relatively small perturbation,
and for the 1.4~ns bunch spacing its effect is small, whereas
for the 2.8~ns spacing it is not.
The reason is that in the former case the beam sits on the half-integer
resonance (which is benign), while in the latter case it sits
on the integer (which is not)\cite{BL}.
As to the effect in a linac, let us distinguish two types of errors:
``systematic random'' and ``purely random'' errors; by the former we
mean errors, random in one structure, that are repeated in all structures
of the linac; by the latter we mean
random also from structure to structure.
We expect the effect of a purely random error, of say, $10^{-4}$
(which we think is achievable) to be similar to a systematic
random error of $10^{-4}/\sqrt{N_a}$.  
$N_{a}=140$, 127, 41 in, respectively,
the prelinac, the $e^+$ drive linac, and the $e^-$
booster; therefore
the appropriate abscissas in the figure become
 .8, .9, and $1.6\times10^{-5}$. At these points, for the 2.8~ns spacing, 
we see that $S_{rms0}$ is
only a factor $2\pm1$, $2\pm1$, $3\pm2$ times
larger than the error-free result. 

\vspace*{-8mm}
\section{Tolerances}
To obtain tolerances we performed particle tracking using
LIAR\cite{LIAR} and compare the results with the analytical
formulas given in Sec. 2. 
We take $\delta\epsilon_t=10\%$ as acceptable.
For BBU the tightest tolerance is
for the $e^+$ booster, where $r_t$ 
is 3.8 (2.2) analytically, 5.5 (3.0) numerically,
for $\Delta t=2.8$ (1.4)~ns.
For misalignments the tightest tolerance is
for the prelinac, where $x_{at}$ 
is 2.9 (4.6)~mm analytically, 3.2 (4.8)~mm numerically.
(For the other machines these tolerances are $\gtrsim10$ times looser.)
Purely random machining errors, 
equivalent to $10^{-4}$ frequency errors,
will tighten these results by 50-100\%, but they are still very loose.

Finally, what is the random, 
cell-to-cell misalignment tolerance? Performing the
perturbation calculation mentioned earlier for 1000 different
random structures,
we find that $S_{rms}=.27\pm.12$ ($.032\pm.003$)
~MV/nC/m$^2$ for $\Delta t=2.8$ (1.4)~ns.
We again see the effect of the integer resonance on the 2.8~ns option result. 
For the prelinac the cell-to-cell misalignment tolerance becomes
40 (600)~$\mu$m for the 2.8 (1.4)~ns configuration.

 We thank T.~Raubenheimer and attendees of the
NLC linac meetings at SLAC for comments and suggestions.


\vspace*{-2mm}
\begin{thebibliography}{9}
\small
\vspace*{-1.0mm}
\bibitem{zdr}
NLC ZDR Design Report, SLAC Report 474, 589 (1996).
\vspace{-1.5mm}
\bibitem{BL}
K. Bane and Z. Li, SLAC-LCC-043, July 2000.
\vspace{-1.5mm}
\bibitem{chao}
    A.\ Chao, ``Physics of Collective Instabilities
    in High-Energy\vspace{-.7mm} Accelerators'', John Wiley \& Sons, New York
    (1993).
\vspace{-1.5mm}
\bibitem{static}
K. Bane, {\it et al}, 
EPAC94, London, England, 1994, p.~1114.
\vspace{-1.5mm}
\bibitem{perturb}
R.~M.~Jones, {\it et al},
PAC99, New York, NY, 1999, p.~3474.
\vspace{-1.5mm}
\bibitem{Gluck}
K. Bane and R. Gluckstern, {\it Part. Accel.}, {\bf 42}, 123 (1994).
\vspace{-1.5mm}
\bibitem{OMEGA2}
X. Zhan, 
PhD Thesis, Stanford University, 1997.
\vspace{-1.5mm}
\bibitem{LIAR}
R. Assmann, {\it et al}, LIAR Manual, SLAC/AP-103, 1997.
\end{thebibliography}
\end{document}